\documentclass[aps]{revtex4}
\usepackage{graphicx}

\newcommand{\cA}{{\cal A}}
\newcommand{\be}{\begin{equation}}
\newcommand{\ee}{\end{equation}}
\newcommand{\ba}{\begin{array}}
\newcommand{\ea}{\end{array}}
\newcommand{\y }{\'\i }

\begin{document}

\pagestyle{plain}

\title{ 
{\bf Symmetry energy coefficients for asymmetric nuclear matter 
 } }
\author{  {\bf F\'abio L. Braghin} \thanks{e-mail: 
braghin@if.usp.br}  \\
{\normalsize 
 Nuclear Theory and Elementary Particle
 Phenomenology Group,}\\
{\normalsize 
Instituto de F\'\i sica, Universidade de 
S\~ao Paulo, C.P. 66.318; CEP 05315-970;  S\~ao Paulo - SP, Brazil. }\\
}
{\it Oral Communication given at Reuni\~ao de Trabalho de F\y sica Nuclear
no Brasil, S\~ao Pedro, SP, Brazil, August 31-September 04, 2002}


\begin{abstract} 
Symmetry energy coefficients of asymmetric nuclear matter
are investigated as the inverse of nuclear matter 
polarizabilities with two different 
approaches. Firstly a general calculation shows they may depend
on the neutron-proton asymmetry itself. 
The choice of particular prescriptions
for the density fluctuations lead to certain isospin (n-p asymmetry)
dependences of the polarizabilities.
Secondly, with Skyrme type interactions, the static 
limit of the dynamical polarizability is investigated
corresponding to the inverse symmetry energy coefficient
which assumes different values at different asymmetries (and 
densities and temperatures). 
The symmetry energy coefficient (in the isovector channel) is found
to increase as n-p asymmetries increase.
The spin symmetry energy coefficient is also briefly investigated.
\end{abstract}

\maketitle

PACS numbers: 21.30.-x, 21.65.+f, 26.50.+x, 26.60.+c

Key-words: Symmetry energy coefficients, nuclear density, n-p asymmetry,
 spin.

{\it Published in Brazilian Journal of Physics, June (2003)}


\section{Introduction}

The (n-p) symmetry energy coefficient and its dependence on the
nuclear density has been extensively studied and this is 
 of relevance, for example,
for the description of macroscopic nuclear properties
as well as for 
proto-neutron and neutron stars.
It represents the 
tendency of nuclear forces to have
greater binding energies (E/A) for symmetric systems - equal number of
protons and neutrons.
It contributes as a  coefficient for the squared 
neutron-proton asymmetry in usual macroscopic mass formula,
$E/A = H_0(A,Z) + a_{\tau}(N-Z)^2/A^2,$
where $H_0$  does not depend on the asymmetry,
Z, N and A are the proton, neutron and mass numbers respectively. 
Other powers of the asymmetry (proportional to $(N-Z)^n$
for $n \neq 2$ \cite{JACO}) 
are usually expected to be less relevant
for the equation of state (EOS) of nuclear
matter  based on such parameterizations \cite{LKLB,MONI}.
The same kind of parameterization is considered for nuclear matter 
where instead of nucleon numbers one has to deal with densities.
$a_{\tau}$ is also the parameter
which measures the response of the system to a perturbation which
tends to separate protons from neutrons. 
It is
given by the static polarizability of the system 
which also may depend on the asymmetry of the medium. 
This point has been developped and emphasized recently \cite{FLB99,ISOSYMEN}.
The
 spin symmetry energy coefficient of nuclear matter may also be 
defined, $a_{\sigma}$, 
representing the cost in energy to make the system spin-asymmetric
(and eventually polarized nuclear matter).
The spin channel is relevant for the study of the neutrino interaction
with matter because it couples with axial vector current together with 
the scalar channel \cite{SAWYER,ESPANHOIS,REDDY}. 
A suppression of the spin susceptibility 
(in this work we will be dealing rather with its inverse)
leads to the suppression of Gamow Teller transitions which are of interest 
for the supernovae mechanism \cite{REDDY} and 
eventually to instabilities associated to 
ferromagnetic polarized states \cite{KUTSCHERA,REDDY}.
In this work we articulate and extend the ideas developped previously
for the dependence of symmetry energy coefficients on 
neutron-proton asymmetry. For this we use 
a calculation for the static  polarizabilities -
proportional to the inverse of the
symmetry coefficients in asymmetric
matter - which was done using Skyrme effective forces 
in \cite{ISOSYMEN,FLB99}.

\section{Generalized Symmetry Energy Coefficients}

When considering a small amplitude external perturbation 
$\epsilon$, the medium polarizability is defined as the ratio of the density 
fluctuation ($\beta = \delta \rho_n - \delta \rho_p$) 
to the amplitude of the external perturbation
and it can be written as \cite{ISOSYMEN,BVA}:
\be \label{1bs} \ba{ll}
\displaystyle{ \Pi^{s,t} \equiv \frac{\beta}{\epsilon}
= - \frac{\rho}{2 \cA_{s,t} (b,\beta) },
}
\ea
\ee
where $\cA_{s,t}$  is the neutron-proton (isovector) 
symmetry coefficient ($s=0,t=1$ - spin, isospin) and 
$b = \rho_n/\rho_p - 1$ is an asymmetry coefficient.
For the other channels (different $(s,t)$) one may define
different symmetry coefficients.
Note that $\cA_{s,t}$ is a function of $b$ and $\beta$ and these 
parameters may be related, as argued below. 
The occurrence of these functional dependences of 
$\cA_{s,t}$ can be found just by 
the first stability condition with respect to the 
(density) fluctuation 
from which one defines the polarizability (\ref{1bs}):
$ d H/d \beta = 0.$ 

\subsection{ Isospin dependence of $\cA_{s,t}$}

We consider
$\cA_{s,t}$ to be a function of the density fluctuation $\beta$.
Although $\beta$ is not the 
explicit n-p asymmetry (given by $b$) we will consider that
it depends on it (as it was also argued in \cite{ISOSYMEN}). 
We consider these parameters are related to each other
and therefore we will write ${\cA} = \cA (\beta)$ shortly.
In \cite{FLB99} two different prescriptions were discussed for
$\beta$ in the calculation of the response function of asymmetric 
nuclear matter. 
We have used (and it was shown to be the more reasonable prescription) 
the  one which leads to the following relation
between  the fluctuation $\beta$ and the 
explicit asymmetry ($b$): 
\be \label{relac} \ba{ll}
\beta = \delta \rho_n \left( \frac{2+b}{1+b} \right),
\ea
\ee
Where $\delta \rho_n$ is the neutron density fluctuation. 
In the n-p symmetric limit $\beta = 2 \delta \rho_n$ and in another limit, 
in neutron matter, $\beta = \delta \rho_n$.
The above prescription (expression (\ref{relac})) 
is based on the assumption that
the density fluctuations are proportional to the respective 
density of neutrons and protons, i.e., 
$\delta \rho_n/ \beta = \rho_n / \rho$, being $\rho$ the total 
density.
In spite of being rather well suited for the isovector channel, 
this kind of assumption can be considered as a starting point 
for the 
other channels (spin, scalar) in asymmetric nuclear matter.
Prescription (\ref{relac})
is therefore model-dependent and different choices for it yield other
 forms for the
the (asymmetric) static screening functions.
The  dynamic response functions are less sensitive to this
prescription \cite{FLB99}.

From the solution of the polarizability (\ref{1bs}) we calculate
the first derivative with relation to $b$:
\be \label{DERIV1} \ba{ll}
\displaystyle{ \frac{d \beta}{d b} = \frac{\epsilon \rho}{2 \cA^2_{0,1} } 
\frac{d \cA^{0,1} }{d b} = - \frac{\beta}{\cA_{0,1}} \frac{d \cA^{0,1}}{d b}. }
\ea
\ee
Another expression can be obtained from the relation between
$b$ and $\beta$ of (\ref{relac}). It yields:
\be \label{relder} \ba{ll}
\displaystyle{ \frac{d \beta}{d b} = - \frac{\beta}{(2+ b)(1+b)}.}
\ea
\ee
Equating these two last equations we obtain:
\be \label{relder2} \ba{ll}
\displaystyle{  - \cA^{0,1} \frac{\beta}{(2+ b)(1+b)} = 
- \beta \frac{d \cA^{0,1}}{d b}
,}
\ea
\ee
From which it is possible to  derive the following relation 
between the isospin s.e.c. and the n-p asymmetry
\cite{ISOSYMEN}:
\be \label{isosb} \ba{ll} 
\displaystyle{ {\cal A}^{0,1} = {\cal A}^{0,1}_{sym} \frac{ 2 + 2 b}{2 + b} 
.}
\ea
\ee
In this expression  $\cA_{sym} = a_{\tau} \simeq 30 MeV$ is the 
s.e.c. of symmetric nuclear matter ($b=0$).
For $b= 2$ (neutron density three times larger than the
proton density) we obtain $\cA = 1.5 \cA_{sym}$. In the
limit of neutron matter 
${\cal A} (b \to \infty) = 2 {\cal A}_{sym}$.

Other prescriptions for the density fluctuation, leading to different
dependences on the n-p asymmetry ($b$),
can be given by:
\be \label{relac2} \ba{ll}
\displaystyle{ 1) \beta = const\;\; \to \;\; {\cal A}_{0,1}=const, } \\
\displaystyle{
2) \beta = \delta \rho_n \frac{2+b^m}{1+b^m} \;\; \to  
\;\; {\cal A}_{0,1}= {\cal A}_{sym} \frac{ 2 + 2 b^m}{2 + b^m}   
(m>1 \;\;\;\; integer)
.}
\ea
\ee
The first of these alternative prescriptions leads to
a constant symmetry energy coefficient ${\cal A}= a_{\tau}$. 
However, it corresponds to
$\delta \rho_n = \delta \rho_p$, independently of the asymmetry of the
medium. 
This does not seems to be reasonable for example because in the
limit of neutron matter there would be no proton density. 
Furthermore this prescription has been used in the dynamical response 
function  yielding seemingly non physical results \cite{FLB99}.
The second of prescriptions (\ref{relac2}) is a very general one
valid for arbitrary integer numbers $m$ and was considered for the
sake of simple algebraic calculation.
Probably $m$ should be not large because it would lead to a too much
 strong (stiff) dependence on $b$.

Another assumption for deriving expressions (\ref{isosb}) and 
(\ref{relac2}) was that 
$\rho$ is independent of $b$. 
This would be non trivial if one considers a complete self consistent
calculation with the equation of state of a proto-neutron star, for
example.

\section{ Polarizabilities with Skyrme forces }

A nearly exact expression for the dynamical polarizability 
of a non relativistic hot asymmetric nuclear 
matter  at variable densities was derived with Skyrme 
interactions in \cite{FLB99}.

The general static screening function $A_{s,t}$ 
in asymmetric nuclear matter at finite temperature was explicitely
written in \cite{ISOSYMEN}.
(The coefficient $b$ is related to a frequently used  
asymmetry coefficient:
$\alpha = (2\rho_{0n}-\rho_0)/\rho_0$, 
by the expression: $b = 2\alpha / (1- \alpha ) $.)


\subsection{Results for Skyrme interactions}

The  Skyrme interactions used are:
SLyb \cite{CHABANAT}, SkSC4, SkSC6 and SkSC10 \cite{DUTOABO,ONSIPP}. 
In Figure 1 we show the inverse of the static polarizability 
(generalized symmetry energy coefficient) in the isovector channel
for diverse Skyrme forces. 
as a function of the asymmetry parameter $b$. 
For symmetric nuclear matter $b=0$ the different Skyrme forces
yield values between $27$ and $34$ MeV, the usual values adopted in the
litterature. For increasing $b$ the coefficient increases, being the 
slope strongly dependent on the interaction.
We plot one case for dense nuclear matter $\rho = 2 \rho_0$ 
(long dashed-short dashed line) with force SLy(b). We note
that the increase of $A_{0,1}$ is much smaller.
These results were discussed more extensively in \cite{ISOSYMEN}.

In Figure 2 the spin symmetry energy is shown as a function
of $b$ with forces SLy, SkSC4, SkSC6 and SkSC10.
The values for symmetric nuclear matter are very different.
 The common trend is the increase of $A_{1,0}$
with $b$, i.e., at very asymmetric matter the spin interaction tends to 
become more repulsive. 
However the particular behavior of the spin s.e.c. with $b$ is different for
each effective force at a given density.
We can compare our results to the ratio of spin susceptibility  
of interacting {\bf neutron matter} 
to the non interacting Fermi gas obtained by
Fantoni, Sarsa and Schmidt \cite{FSS2001} by means of the 
auxiliary field diffusion Monte Carlo method. 
This ratio is proportional to the polarizability as obtained in expression
(\ref{1bs}) and therefore inversely proportional to the spin
symmetry coefficient $A_{1,0}$. 
First of all we note that, in most cases,
the values they find are 
all positive for the range of densities 
considered by them, from $0.75 \rho_0$
up to $2.5 \rho_0$ (in our calculation the total density was kept 
constant and equal to the saturation density $\rho_0$). 
The slope seems to be nearly the same as that we obtain
for low values of the n-p asymmetry. 
Consequently they may obtain
instabilities for higher density neutron matter whereas  we do not 
observe this result in our calculations with Skyrme forces at
the saturation density 
(the comparison is meaningful for neutron matter: $b$ very large).
A further comparison at different total nuclear densities
is to be shown elsewhere \cite{ISOSYMEN}.

\section{ Summary and Conclusions}

Summarizing, the dependence of the s.e.c. on the n-p asymmetry
was studied extending the results of ref. \cite{ISOSYMEN}.
Different prescriptions for the density fluctuations (expressions
(\ref{relac},\ref{relac2})) in asymmetric nuclear matter lead to different 
dependences of the symmetry energy coefficients on the asymmetry
parameter $b$.
The 
n-p asymmetry dependence of the s.e.c. in the different channels
was analyzed for
different Skyrme forces. They may yield very different
behaviors including the possibility  (or not) 
of nuclear matter to undergo phase transitions.
These forces can describe different behaviors
of the symmetry energy coefficients.
Therefore, in principle, different values can be expected  for the 
(bulk) symmetry energy coefficients in  asymmetric nuclear matter
with different n-p asymmetries. 
It would be interesting to apply the results obtained here and in 
\cite{ISOSYMEN} in studies of the equation of state of 
asymmetric nuclear matter, trying to extract 
experimental constraints for ${\cal A_{s,t}}$, in particular
in the isospin channel ($s=0, t=1$).
(For works which can be related with the present ideas see, for example, 
these proceedings and \cite{BAOANLI}).
Although we have dealt with neutron-proton density asymmetry
we can expect that the same kind of ideas can be applied for
the neutron-proton number asymmetry used in mass formulas
for finite nuclei \cite{MONI}.
In the spin channel  it is possible to expect
spin polarized asymmetric matter 
yielding magnetic fields in neutron stars, as discussed
in \cite{KUTSCHERA}.
However with the increase of $b$ we find that
the spin interaction may be rather repulsive, hindering
this magnetization effect with the use of these Skyrme forces
for the value of density analyzed here ($\rho_0$).
In this work we do not include the simultaneous study of the
density dependence of the symmetry coefficients.

\vskip 0.3cm
\noindent {\Large {\bf Acknowledgement}}

This work was supported by FAPESP, Brazil.

\vspace{2cm}

\noindent {\Large {\bf  Figure captions}}

\vskip 0.5cm

{\bf Figure 1}  Neutron-proton symmetry energy coefficient 
$A_{0,1}= \rho/(2 \Pi_R^{0,1})$ as a function of the
asymmetry coefficient $b$ for several Skyrme forces: 
dotted line (SGII), thick dotted (SkSC4), long-short dashed (SLy(b)),
crosses (SkSC6), thick long-short dashed (SLy(b) at twice $\rho_0$),
medium thick dotted (SkSC10).

{\bf Figure 2}  Spin symmetry energy coefficient 
$A_{1,0}= \rho/(2 \Pi_R^{1,0})$ as a function of the asymmetry
coefficient $b$  for the different forces:
 thick dotted (SkSC4), long-short dashed (SLy(b)),
medium thick dotted (SkSC6), 
 thick solid line (SkSC10).

\begin{figure}[htb]
\includegraphics[width=9cm]{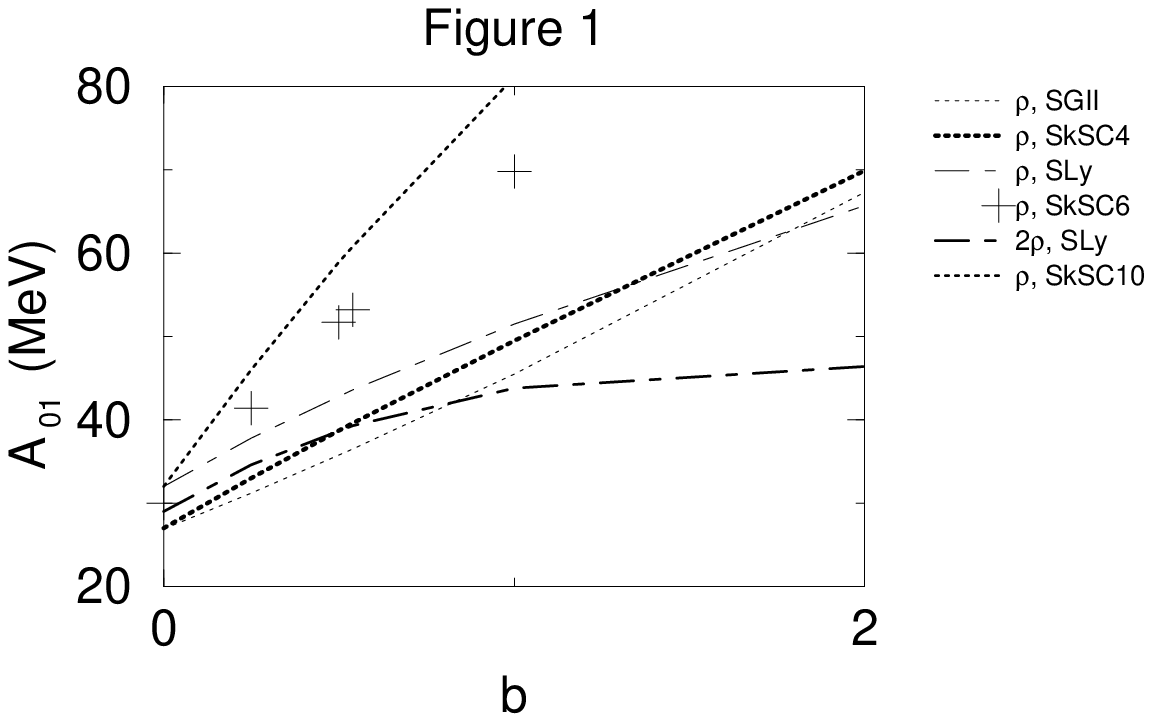}
\end{figure}

\begin{figure}[htb]
\includegraphics[width=9cm]{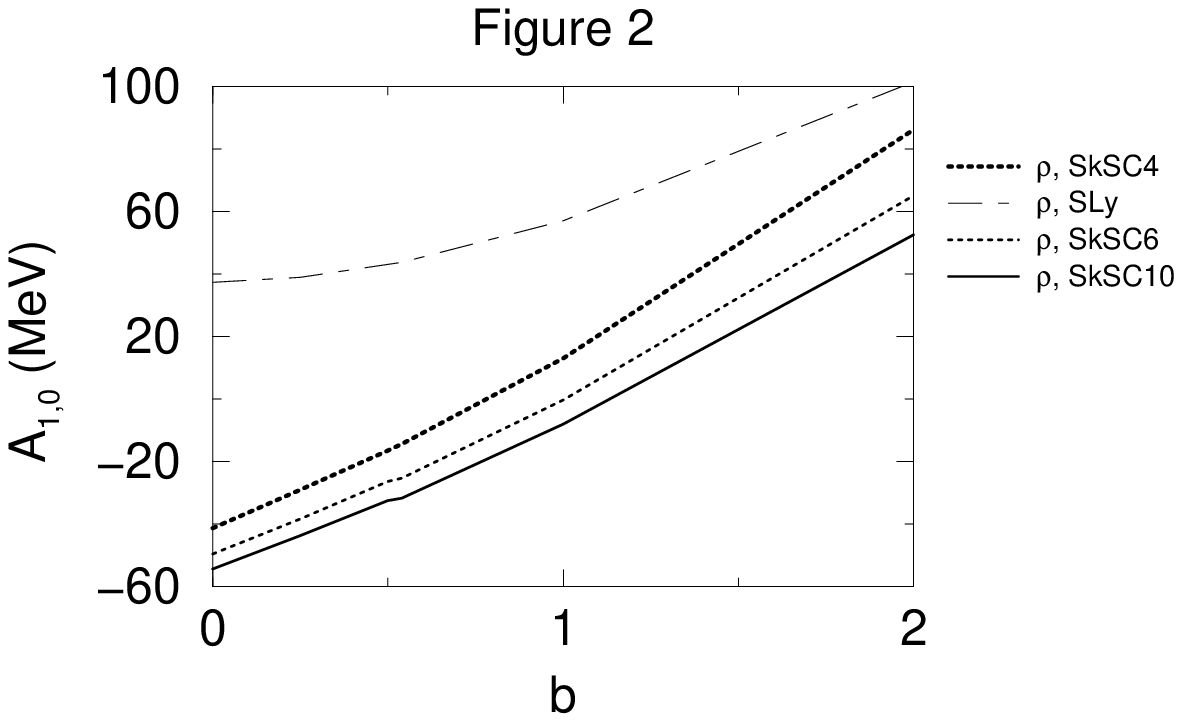}
\end{figure}


\vskip 0.2cm



\begin{thebibliography}{ll}
\bibitem{JACO} J. J\"anecke and E. Comay, Nucl. Phys. {\bf A 436} 
(1985) 108.
\bibitem{LKLB} C.-H. Lee {\it et al}, Phys. Rev. {\bf C 57} (1998) 3488.
\bibitem{MONI} 
P. Moeller, J.R. Nix, W.D. Myers and W.J. Swiatecki,
At. Data Nucl. Data Tables {\bf 59}, 185 (1995);
Y. Aboussir, J.M. Pearson, A.K. Dutta and F. Tondeur,
At.  Data Nucl. Data Tables {\bf 61}, 127 (1995).
W.D. Myers and W.J. Swiatecki, Nucl. Phys. {\bf A 601}, 141 (1996).

\bibitem{FLB99} F.L. Braghin, Phys. Lett. {\bf B 446}, (1999) 1;
 Nucl. Phys. {\bf A 665}, (2000) 13.

\bibitem{ISOSYMEN} F.L. Braghin, 
Nuc. Phys. {\bf A 696} (2001) 413; F.L.B., 
{\it Erratum} Nuc. Phys. {\bf A} to be published.
F.L.B. {\it submitted to publication}.


\bibitem{SAWYER} R.F. Sawyer, Phys. Rev. {\bf C 11}, 2740 (1975), 
N. Iwamoto and C.J. Pethick, Phys. Rev. {\bf D 25}, 313 (1982).
\bibitem{ESPANHOIS} J. Navarro, E.S. Hernandez, D. Vautherin, 
Phys. Rev. {\bf C 60} (04)5801 (1999).
\bibitem{REDDY} S. Reddy, M. Prakash, J.M. Lattimer, J.A. Pons,
  Phys. Rev. {\bf C 59}, (1999) 2888.

\bibitem{KUTSCHERA} M. Kutschera and W. W\'ojcik, Phys. Lett. {\bf B 223},
11 (1989)

\bibitem{BVA} F.L. Braghin, D. Vautherin and A. Abada, Phys. Rev. {\bf C
52},  (1995) 2504.

\bibitem{CHABANAT} E. Chabanat {\it et al},  Nucl. Phys. {\bf A 627}, 
 (1997) 710.
\bibitem{DUTOABO} A.D. Dutta {\it et al}, Nucl. Phys. {\bf A 458},
77 (1986); F. Tondeur {\it et al}, Nucl. Phys. {\bf A 470}, 93 (1987),
Y. Aboussir {\it et al}, Nucl. Phys. {\bf A 549}, 155 (1992).
\bibitem{ONSIPP} M. Onsi, H. Przysiezniak and J.M. Pearson,
Phys. Rev. {\bf C 50}, (1994) 460.

\bibitem{FSS2001} S. Fantoni, A. Sarsa, K.E. Schmidt, Phys. Rev. Lett. {\bf 87},
181101 (2001)

\bibitem{BAOANLI} Bao-An Li, Nucl. Phys. {\bf A 708} (2002) 365.

\end{thebibliography}
\end{document}